\documentclass{article}

\usepackage{PRIMEarxiv}

\usepackage[utf8]{inputenc} 
\usepackage[T1]{fontenc}    
\usepackage{hyperref}       
\usepackage{url}            
\usepackage{booktabs}       
\usepackage{amsfonts}       
\usepackage{nicefrac}       
\usepackage{microtype}      
\usepackage{lipsum}
\usepackage{fancyhdr}       
\usepackage{graphicx}       
\graphicspath{{media/}}     

\usepackage{amsmath,amsfonts,amssymb}
\usepackage{graphicx}
\usepackage{setspace}
\usepackage{tocloft}
\usepackage{bm}

\usepackage{lineno}
\usepackage{algorithm}
\usepackage{algpseudocode}
\usepackage[title]{appendix}
\usepackage{siunitx}

\title{High-fidelity near-diffraction-limited projection through scattering with reference-less transmission matrix
\\ \thanks{\textit{\underline{Citation}}: 
\\ \textbf{{\dag}Contacts: qingyang@zju.edu.cn, wenzhong@zju.edu.cn}} 
}

\author{ {Jingshan Zhong} \\
	Research Center for Humanoid Sensing\\
    Zhejiang Lab\\
    Hangzhou 311100, China\\
	\texttt{zhongjingshan@hotmail.com} \\
	\And
	Quanzhi Li, {Zhong Wen{\dag}, Qilin Deng, Haonan Zhang, Weizheng Jin and Qing Yang{\dag}} \\
	State Key Laboratory of Extreme Photonics and Instrumentation\\
    College of Optical Science and Engineering\\
    International Research Center for Advanced Photonics\\
    Zhejiang University, Hangzhou 310027, China\\
	\texttt{\{qingyang,wenzhong\}@zju.edu.cn} \\
}

\begin{document}
\maketitle

\begin{abstract}

Image projection through scattering media has applications ranging from light delivery through multimode fiber to near-eye displays. Conventional methods utilize the transmission matrix (TM) measured by interfering with a reference beam. However, it is noise-sensitive, often resulting in artifacts that degrade the projection quality. Here we propose to characterize the scattering by computationally retrieving TM from intensity-only measurements and solve the projection problem formulated with the retrieved TM by optimization. We experimentally validate the proposed method by projecting through a multimode fiber. Compared to the conventional methods, it projects improved-quality images with resolution near to the diffraction limit, and simplifies the experimental setup by eliminating the reference. It paves the way for applications of high-quality near-diffraction-limited projection through scattering.
\end{abstract}


\section{Introduction}

\begin{figure}[tb]
	\centering
	\includegraphics[width=0.8\linewidth]{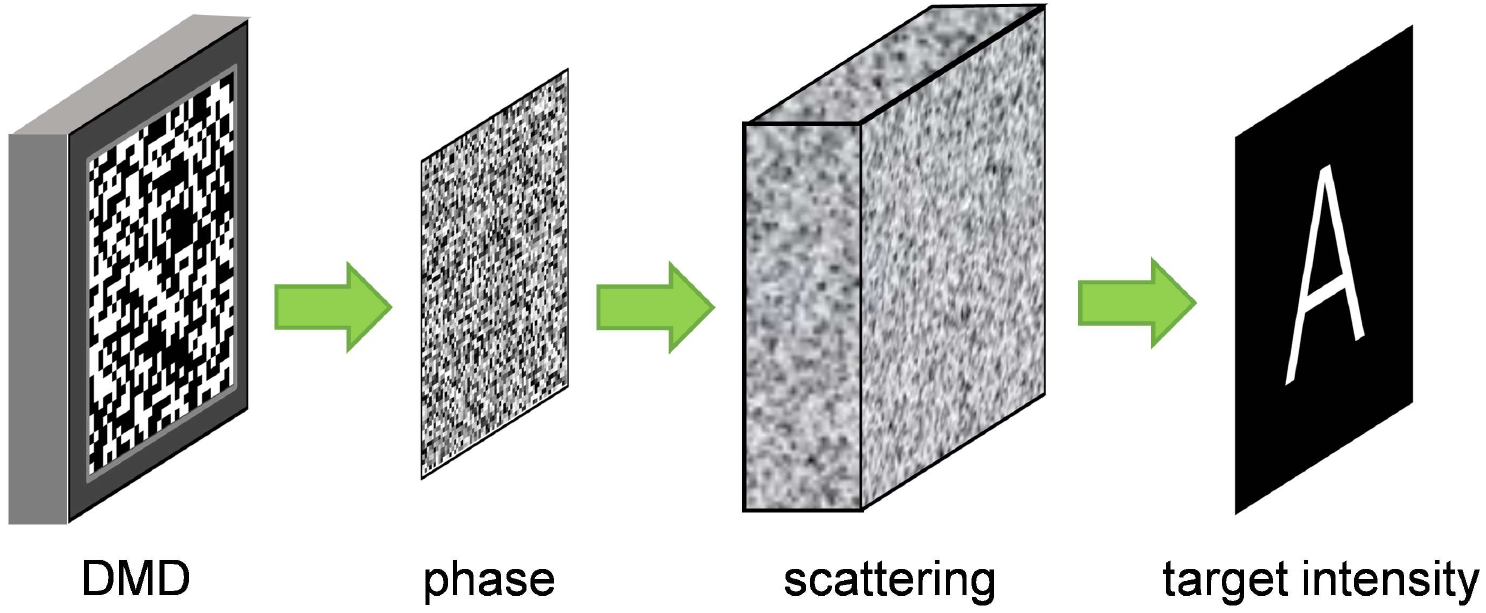} 
	\caption{Image projection through scattering. The phase of the incident complex field is modulated by using a DMD or spatial light modulator (SLM). The complex field  transmitted through the scattering media is manipulated according to the distribution of the target intensity.}
	\label{fig:projection}
    \vspace{-0.4cm}
\end{figure}

Light scatters randomly when it travels through scattering media. Through modulation of the incident light, the transmitted light could be manipulated into a target intensity distribution (fig.~\ref{fig:projection}). It allows to utilize scattering media to overcome the limitations of conventional optical systems. For example, projecting light through multimode fiber (MMF) enables applications such as micro-endoscopic imaging~\cite{ploschner2015seeing, wen2022single}, optical tweezers through MMF~\cite{leite2018three}, structured light delivery in opto-genetics~\cite{ronzitti2017recent} and MMF lithography~\cite{konstantinou2023improved}. Projection through scattering mask breaks the space-bandwith product limit of the spatial light modulator, addressing the tradeoff between eyebox and field of view (FOV) in the holographic display~\cite{yu2017ultrahigh,kuo2020high}. High-fidelity image projection is essential to these applications, which require stimulating neuronal activity, controlling nano-particles, exposing photoresist in lithography, and improving the visual quality of displays. Therefore, it is crucial to develop methods to precisely project light through scattering media.

The projection involves characterizing the scattering media and developing inverse algorithms to calculate the modulations of the incident light. The transmission matrix (TM) is commonly used to model the scattering property. The interference-based TM calibration method~\cite{vcivzmar2011shaping} measures the transmitted complex fields (including both amplitudes and phases) through interference with an external reference beam. With the measured TM, the modulation patterns for light projection are obtained through inversion~\cite{xu2017focusing}, back-propagation~\cite{ploschner2015multimode}, or iterative forward-backward propagation (Gerchberg-Saxton)~\cite{bianchi2012multi}. In addition, the optimization-based methods~\cite{flaes2021time,yu2024high} solve the modulations by minimizing the error between the intensity predicted by the measured TM and the target intensity. However, the interference in the TM calibration is noise-sensitive due to phase instability and mechanical movements, leading to speckles and artifacts in the projected image. Instead of using an external reference beam, the TM calibration can allocate a portion of the modulation modes as the internal reference beam~\cite{popoff2010measuring}. It has potential issues of missing points due to the dark regions in the internal reference. Several projection methods~\cite{yu2017ultrahigh, yu2023ultrahigh} employ this TM calibration method in three-dimensional display. However, only selected focal points were displayed to demonstrate the potential for expanding eyebox and FOV with scattering. As an alternative to TM, neural networks have been proposed to model scattering by learning from the input-output pairs of the scattering media~\cite{rahmani2020actor,huang2023projecting}. However, the projected image using neural networks still suffers from speckles, probably due to discrepancies between the neural network model and physical reality. Therefore, high-fidelity near-diffraction-limited grayscale projection through scattering remains challenging, requiring co-design of scattering calibration and projection algorithm.

\section{Method}
\label{sec:headings}

\begin{figure}[htbp]
	\centering
	\includegraphics[width=0.8\linewidth]{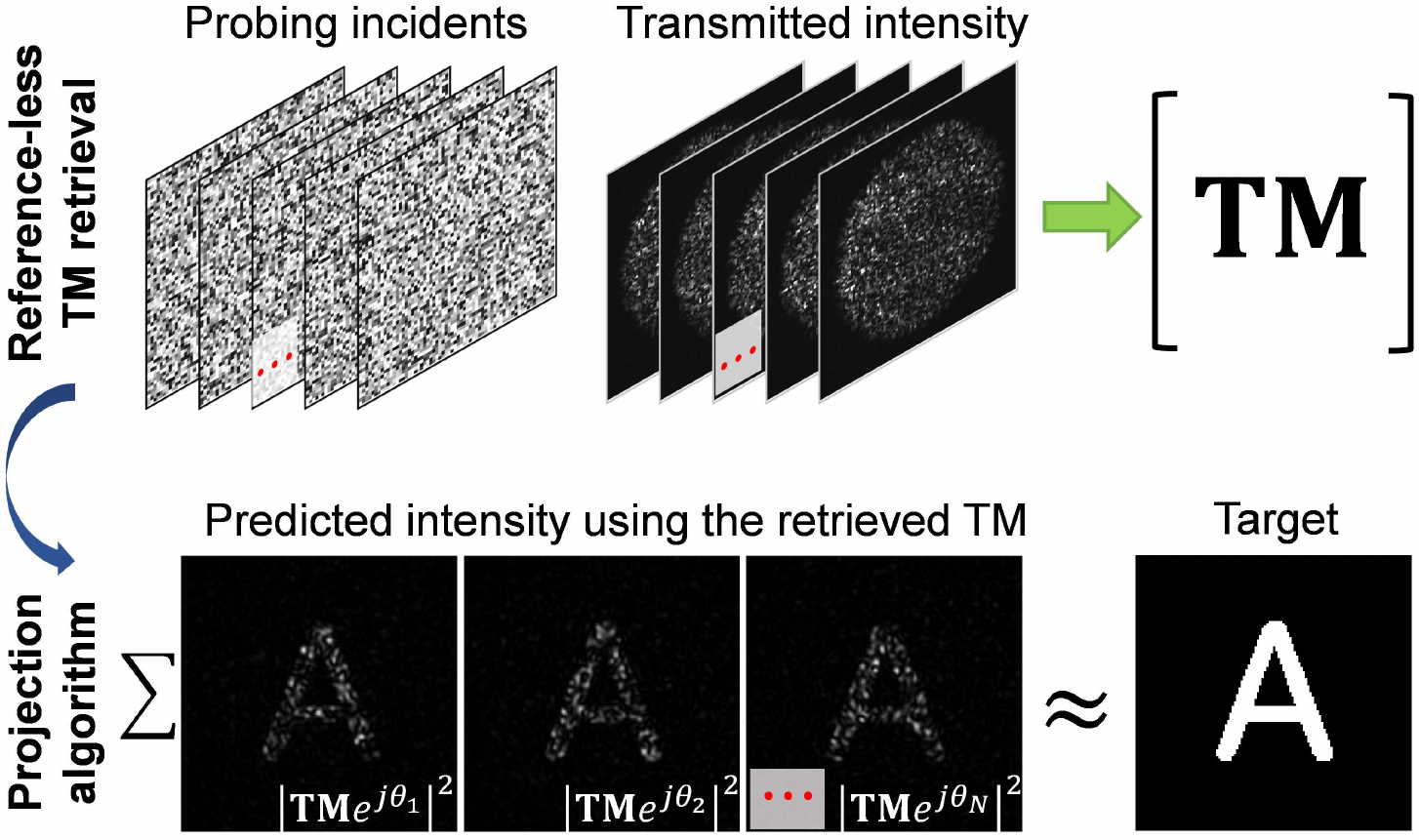} 
	\caption{ The proposed method for projection through scattering. The TM of the scattering is recovered from input-output pairs of probing incident complex fields and the transmitted intensity. The projection algorithm solves the incident modulations to best-fit the predicted intensity and the target image. }
	\label{fig:scheme}
\end{figure}

In this work, we propose to calibrate the scattering media by reference-less TM retrieval and design the projection algorithm with the retrieved TM (Fig.~\ref{fig:scheme}). The reference-less TM retrieval reconstructs the TM computationally from the probing incident complex fields and the intensity of the transmitted complex fields. By removing interference with a reference beam, it gains the advantages of simple experimental setup and robustness. In the projection algorithm, the retrieved TM is used to predict the projected intensity. The algorithm solves the modulations of the incident complex fields to achieve best-fit between the predicted intensity and the target intensity.

\begin{figure}[htbp]
	\centering
	\includegraphics[width=0.60\linewidth]{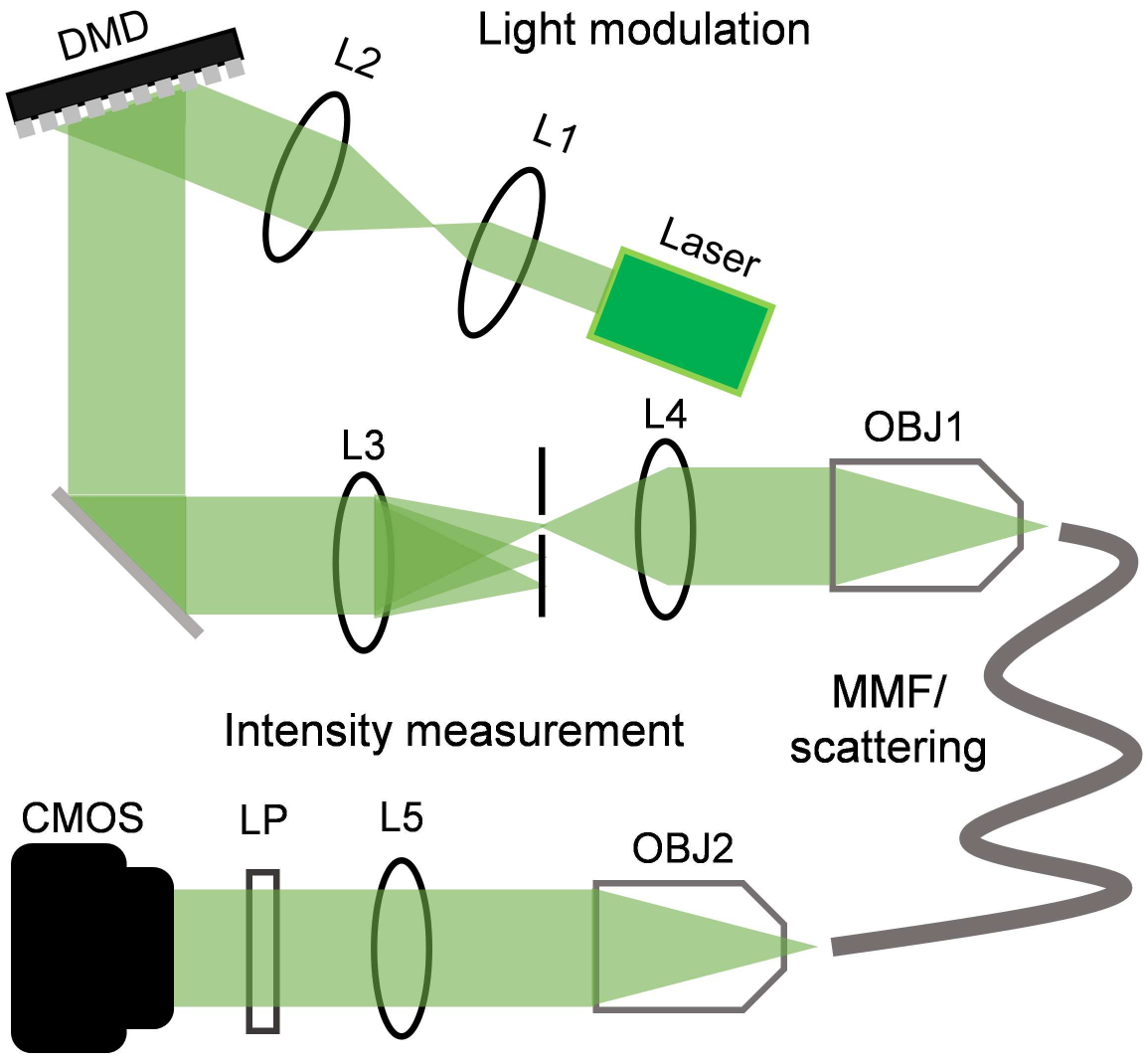} 
	\caption{Experimental setup. The light modulation module modulates the phase of the incident light with a DMD. The intensity measurement module measures the intensity of the transmitted complex fields with a magnifying system. L1-L5: lens, OBJ: objective lens, LP: linear polarizer, CMOS: camera. }
	\label{fig:setup}
\end{figure}

We validate the proposed method by projecting images through a MMF (Fig.~\ref{fig:setup}). A laser beam (788  \si{\nm}) is expanded onto the DMD (Vialux 7001). The 4f system (L3 and L4) filters out the first diffraction order of the light reflected by the DMD. The Lee hologram method~\cite{lee1978computer} is applied to modulate the phase of the incident light. The objective OBJ1 (20X, NA 0.4) focuses the modulated light into the proximal end of the MMF (Optran Ultra WFGE 100/110/125P, NA 0.37 and diameter 105 \si{\um}). Another 4f system, comprised of OBJ2 (20X, NA 0.4) and L5 , magnifies the transmitted complex field, and the camera (Balser acA720-520 \si{\um}) records its intensity. The experimental setup can be adapted to projecting image thorough other scattering media, such as diffuser or ZnO layers.

The forward model of the optical system relates the phase modulation of the incident complex field with the intensity at the camera. It can be expressed with the TM,
\begin{align}
 {\bf{I}}=  {\left| {\bf{T}} e^{j\pmb{\theta}}  \right|}^2,
 \label{eq：physicalmodel}
\end{align}
where $e^{j\pmb{\theta}}$ is the vectorized version of the modulated incident complex field, $\bf{I}$ is the intensity measurement of the transmitted complex field, $\bf{T}$ is the TM, and $\left| \cdot \right| $ takes the absolute square of the inside vector. The TM linearly relates the incident and transmitted complex fields and the intensity at the camera takes absolute square of the transmitted complex field. 

The proposed method learns the TM for projection from input-output pairs of the probing modulation of the incident and the corresponding intensity measurement. We adopt a fast reference-less TM retrieval method~\cite{TM_Zhong2023} which exploits the efficiency of FFT by coding the probing complex fields with modified Fourier transform matrix. It obtains an estimated TM, ${\bf{T}}_{rl}$, which best-fits the intensity measurements and the intensity predicted by ${\bf{T}}_{rl}$.

The TM learned from the reference-less retrieval is used to predict the projected image. The projected image is modelled as the combined outcome of several modulations,
\begin{align}
{\bf{I}}_p = \sum \limits_{m} {\left| {\bf{T}}_{rl} e^{j\pmb{\theta}_m} \right|}^2,
\label{eq:forwardmodel}
\end{align}
where ${\bf{I}}_p$ is the prediction of the projected image, ${\pmb{\theta}}_m$ are the phase modulations of the incident complex field, $m=1, \cdots, M$, and $M$ is the total number of phase modulations used to project the image. Here ${\left| {\bf{T}}_{rl} e^{j\pmb{\theta}_m} \right|}^2$ is the output intensity predicted by the retrieved ${\bf{T}}_{rl}$ when the incident complex field is modulated with $e^{j\pmb{\theta}_m}$. The final projected image is a summation of the $M$ consecutive output intensity images. Employing multiple modulations allows more degree of freedoms to project the image, improving the ability to control the light distribution. It brings the cost of reducing the display frame rate. However, it can be mitigated by the fact that the commercial DMD usually has a frame rate exceeding 23kHz. 

\begin{figure*}[ht]
	\centering
	\includegraphics[width=\linewidth]{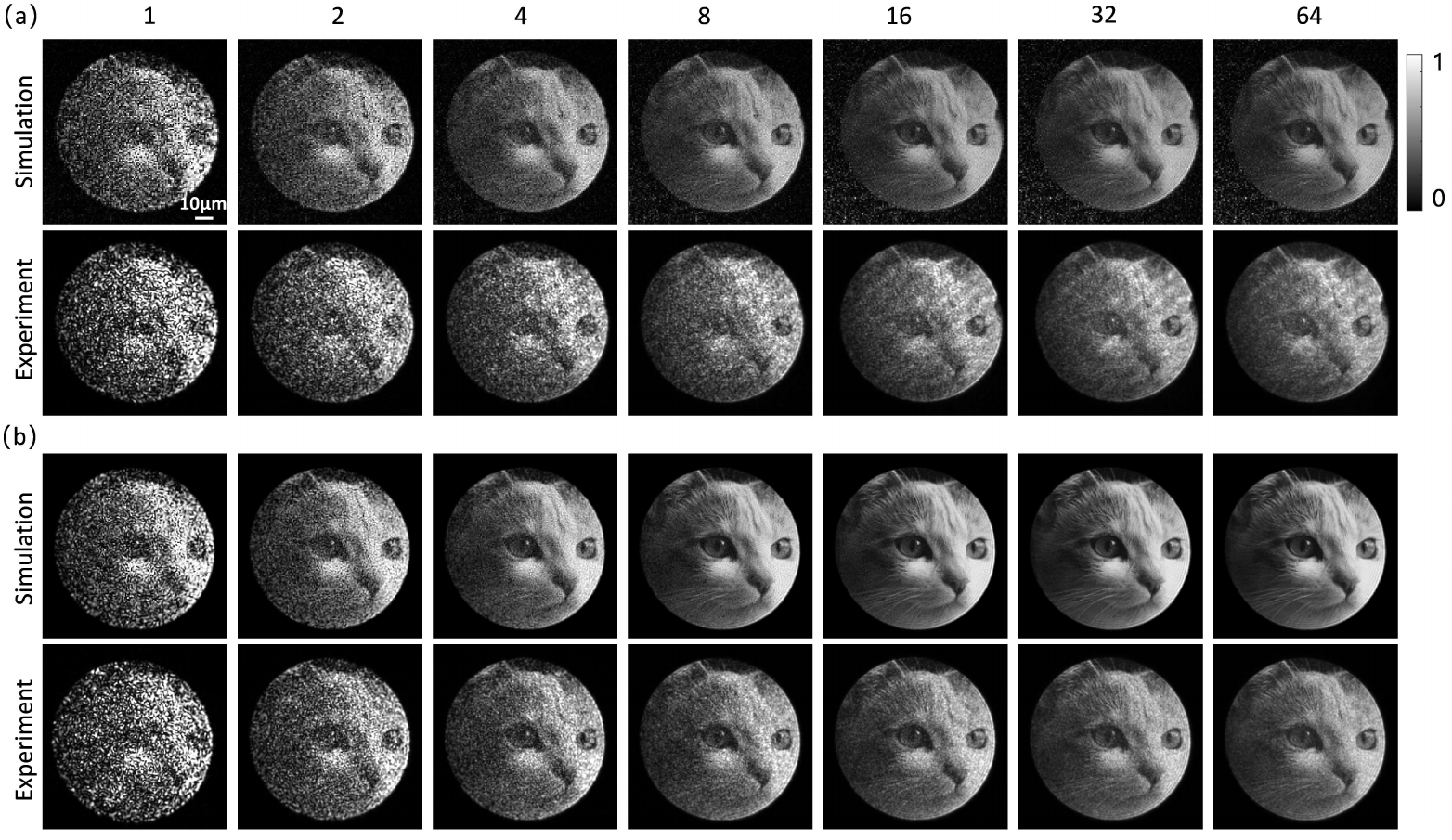} 
	\caption{Comparison of the projected images by the proposed method and the conventional method using the TM measured by the off-axis holography method. (a) The simulated and projected images of the conventional method. The simulated images improve accuracy to the target image as the number of phase patterns increase from 1 to 64. However, the projected images in the experiment suffer from low frequency artifacts. (b) The simulated and projected images of the proposed method. The projected images measured in the experiment match well with the simulated images, showing improved quality.}
	\label{fig:compare}
  \vspace{-0.2cm}
\end{figure*}

The projection algorithm solves the phase modulations which minimize the error between the predicted intensity and the target intensity. The optimization problem is formulated as,
\begin{align}
    \min \limits_{{\pmb{\theta}}_1, {\pmb{\theta}}_2,...,{\pmb{\theta}}_M}  f=\| {\bf{I}} - \alpha \sum \limits_{m} {\left|  {\bf{T}}_{rl} e^{j{\pmb{\theta}}_m} \right|}^2 \|^2_2,
    \label{eq:optcgh}
\end{align}
where the scaling factor $\alpha={\bf{I}}^T{\bf{I}}_p/{{{\bf{I}}_p}^T{\bf{I}}_p}$, $T$ denotes transpose, ${\bf{I}}$ is the target image, and $\|  \cdot \|^2_2$ takes the L2 norm of the inside vector. Instead of being a constant, the scaling factor is a function of the phase modulations which linearly fits ${\bf{I}}$ and ${\bf{I}}^p$. The optimization problem is solved by the algorithm of phase retrieval ~\cite{zhong2016nonlinear}. The first derivative is analytically derived as,
\begin{align}
\frac{\partial  \alpha}{\partial {\bf{I}}_p} = &  \frac{1}{{{\bf{I}}_p}^T{\bf{I}}_p} {\bf{I}}^T - \frac{{{\bf{I}}}^T{\bf{I}}_p}{({{\bf{I}}_p}^T{\bf{I}}_p)^2} {{\bf{I}}_p}^T \\
\frac{\partial f}{\partial {\pmb{\theta}}_m} = & -2 \text{diag}(-je^{-j{\pmb{\theta}}_m}) {\bf{T}}_{rl}^H \text{diag} (  {\bf{T}}_{rl} e^{j{\pmb{\theta}}_m} ) \nonumber \\
& (\alpha + (\frac{\partial  \alpha}{\partial {\bf{I}}_p})^T {\bf{I}}_p^T) ({\bf{I}} - \alpha \sum \limits_{m} {\left|  {\bf{T}}_{rl} e^{j{\pmb{\theta}}_m} \right|}^2).
\label{eq:firstder}
\end{align}
With the cost function (Eq.~\ref{eq:optcgh}) and the first derivative (Eq.~\ref{eq:firstder}), the optimization problem is solved by the second order optimization method such as the limited memory Broyden–Fletcher–Goldfarb-Shanno method~\cite{liu1989limited}.


 Our method utilizes the physical model which relates the modulation to the intensity for both TM retrieval and prediction of the projected image. However, the TM calibrated by interference relates the incident and transmitted complex fields, and its projection prediction suffers from noise sensitivity issues. Moreover, the neural network model exists discrepancy with the physical model in Eq.~\ref{eq：physicalmodel}. In comparison, the proposed method based on the exact physical model guarantees an accurate prediction of the projected intensity. On the other hand, the projection algorithm derives the first derivative analytically. It is more accurate and efficient than the case which numerically approximates the first derivative~\cite{flaes2021time}.

\section{Result}

We validated the proposed projection method by the experimental setup in fig.~\ref{fig:setup}. We first calibrated the TM of the scattering medium for image projection. The TM retrieval method measured 16384 intensity images after sending in the probing incidents designed by the modified Fourier transform matrix. The phase modulation has 2048 modes, and each measured image has 192 by 192 pixels with a pixel size of 0.62 \si{\um}. It recovers a TM of size 36864 by 2048 in 231.8 seconds. Next we implemented the proposed projection algorithm to solve the phase modulations for projection. The incident field was modulated with the each of the obtained modulations, and the corresponding transmitted intensity images were measured. The projected image is obtained by summing over all measured images.

We compared the proposed method with the conventional projection method which uses the TM measured with interference. In the conventional method, the TM was measured using the off-axis holography method with an external reference beamd~\cite{vcivzmar2011shaping}. And the measured TM was used in the projection algorithm to solve the phase modulations for projection. The target image was a gray-scale image of a cat. We conducted projections using both the proposed method and the conventional method, varying the number of phase modulations (1, 2, 4, 8, 16, 32, and 64). Figure \ref{fig:compare} (a) shows the simulated and displayed images of the conventional method. The simulated images are the predicted images numerically obtained using the phase modulations returned by the optimization. As the number of modulations increases, the simulated image gradually approaches the target image. It indicates that the optimization algorithm can project a high-quality image in simulation. However, the experimentally projected images of the conventional method suffer from artifacts, which degrade the projection quality. The mismatch between the simulations and experiments can be attributed to errors in the measured TM. Figure \ref{fig:compare} (b) shows the simulated and projected images of the proposed method using the reference-less retrieved TM. The images in the simulation and experiment match better compared to the conventional method. For the case that the number of modulations is 64, the high-frequency artifacts are almost eliminated. In comparison to the conventional method, the projected images of our method demonstrate improved projection quality with less artifacts.

\begin{figure}[hptb]
	\centering
	\includegraphics[width=0.7\linewidth]{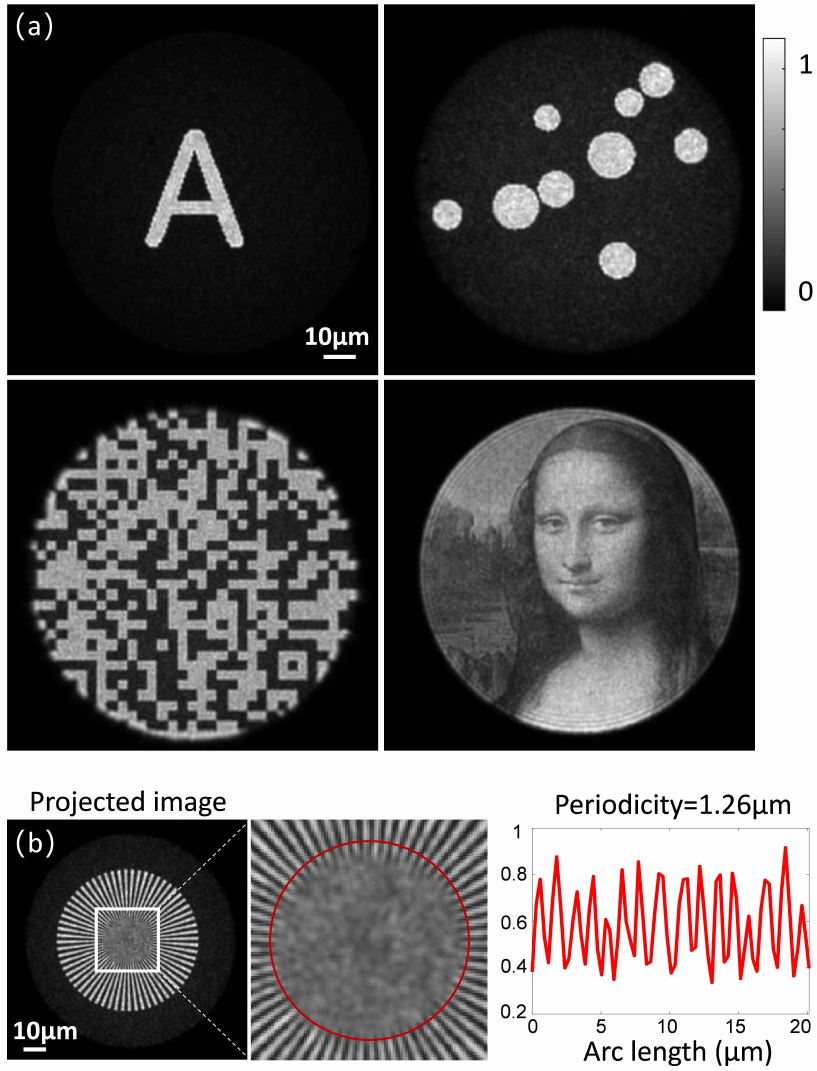} 
	\caption{Additional experimental results and the resolution calibration. (a) The projected images include letter A, random circles, binary code, and Mona Lisa. (b) The resolution of the projection was calibrated by projecting an image of Siemens star. The periodicity of the visually resolvable features (marked as red) in the projected image is 1.26 \si{\um}.  }
	\label{fig:moreresult}
\end{figure}

Figure \ref{fig:moreresult} (a) shows additional projected images by the proposed method. The projected images include cases of letter A, random circles, binary mask, and part of Mona Lisa. It demonstrates the proposed method's capability to project high quality image through scattering. Furthermore, we calibrated the resolution of the projected image by projecting a Siemens star~\cite{horstmeyer2016standardizing}. The resolution of the projected image is 1.26 \si{\um}. It closely approximates the diffraction limit of the MMF, which is 1.06 \si{\um}( $\lambda/2\text{NA}$). The factors causing degradation in the project resolution may include the number of modes in the phase modulation, the alignment of the imaging system, and the number of phase modulations. It demonstrates the method's ability to project high-quality image through scattering with the resolution near the diffraction limit.

\section{Conclusion}

Our work presented a novel method for projecting images through scattering media. The proposed method characterized the scattering media with reference-less TM retrieval and solved the projection problem formulated with the retrieved TM by nonlinear optimization. It was experimentally validated by projecting grayscale images through the MMF. Compared to the conventional method using the TM measured by interference, our method projected images with less artifacts and simplifies the experimental setup by eliminating reference. It projected the high-quality grayscale images with resolution near the diffraction limit. Besides MMF, it can be extended to project through other scattering media, including diffuser, ZnO layers, biological tissues, or a display system with strong aberration. Moreover, our demonstration only used 64 frames in the projection, but the frame rate of DMD is up to 23kHz. Future work could explore fully exploiting the DMD's high frame rate to enhance the projection capability.

\section*{Acknowledgments}
This work was supported by the National Natural Science Foundation of China (62020106002, 62005250, T2293751,
and T2293752), the National Key Basic Research Program of China (2021YFC2401403), and Research Project of Zhejiang Lab (K20231130146TW01).

\bibliographystyle{unsrt}  
\bibliography{references}

\end{document}